\documentclass[aps,prl,floatfix,10pt,superscriptaddress, twocolumn]{revtex4-2}
 
\usepackage[english]{babel}
\usepackage{subcaption}
\usepackage{amsmath}
\usepackage{graphicx}
\usepackage[colorlinks=true, allcolors=blue]{hyperref}

\begin{document}

\title{Design and sensitivity of a 6-axis seismometer for gravitational wave observatories}
\date{\today}

\author{Leonid Prokhorov}
\email{l.g.prokhorov@bham.ac.uk}
\affiliation{Institute for Gravitational Wave Astronomy, School of Physics and Astronomy, University of Birmingham, Birmingham B15 2TT, United Kingdom}

\author{Sam Cooper}
\affiliation{Institute for Gravitational Wave Astronomy, School of Physics and Astronomy, University of Birmingham, Birmingham B15 2TT, United Kingdom}

\author{Amit Singh Ubhi}
\affiliation{Institute for Gravitational Wave Astronomy, School of Physics and Astronomy, University of Birmingham, Birmingham B15 2TT, United Kingdom}

\author{Conor Mow-Lowry}
\affiliation{Institute for Gravitational Wave Astronomy, School of Physics and Astronomy, University of Birmingham, Birmingham B15 2TT, United Kingdom}
\affiliation{Department of Physics and Astronomy, VU Amsterdam, 1081 HV, Amsterdam, The Netherlands}
\affiliation{Dutch National Institute for Subatomic Physics, Nikhef, 1098 XG, Amsterdam, Netherlands}

\author{John Bryant}
\affiliation{Institute for Gravitational Wave Astronomy, School of Physics and Astronomy, University of Birmingham, Birmingham B15 2TT, United Kingdom}

\author{Artemiy Dmitriev}
\affiliation{Institute for Gravitational Wave Astronomy, School of Physics and Astronomy, University of Birmingham, Birmingham B15 2TT, United Kingdom}

\author{Chiara Di Fronzo}
\affiliation{Institute for Gravitational Wave Astronomy, School of Physics and Astronomy, University of Birmingham, Birmingham B15 2TT, United Kingdom}

\author{Christopher J. Collins}
\affiliation{Institute for Gravitational Wave Astronomy, School of Physics and Astronomy, University of Birmingham, Birmingham B15 2TT, United Kingdom}

\author{Alex Gill}
\affiliation{Institute for Gravitational Wave Astronomy, School of Physics and Astronomy, University of Birmingham, Birmingham B15 2TT, United Kingdom}

\author{Alexandra Mitchell}
\affiliation{Department of Physics and Astronomy, VU Amsterdam, 1081 HV, Amsterdam, The Netherlands}
\affiliation{Dutch National Institute for Subatomic Physics, Nikhef, 1098 XG, Amsterdam, Netherlands}

\author{Joscha Heinze}
\affiliation{Institute for Gravitational Wave Astronomy, School of Physics and Astronomy, University of Birmingham, Birmingham B15 2TT, United Kingdom}

\author{Jiri Smetana}
\affiliation{Institute for Gravitational Wave Astronomy, School of Physics and Astronomy, University of Birmingham, Birmingham B15 2TT, United Kingdom}

\author{Tianliang Yan}
\affiliation{Institute for Gravitational Wave Astronomy, School of Physics and Astronomy, University of Birmingham, Birmingham B15 2TT, United Kingdom}

\author{Alan V. Cumming}
\affiliation{Institute for Gravitational Wave Research, School of Physics and Astronomy, University of Glasgow, Glasgow G12 8QQ, United Kingdom}

\author{Giles Hammond}
\affiliation{Institute for Gravitational Wave Research, School of Physics and Astronomy, University of Glasgow, Glasgow G12 8QQ, United Kingdom}

\author{Denis Martynov}
\affiliation{Institute for Gravitational Wave Astronomy, School of Physics and Astronomy, University of Birmingham, Birmingham B15 2TT, United Kingdom}

\begin{abstract}
We present the design, control system, and noise analysis of a 6-axis seismometer comprising a mass suspended by a single fused silica fibre. We utilise custom-made, compact Michelson interferometers for the readout of the mass motion relative to the table and successfully overcome the sensitivity of existing commercial seismometers by over an order of magnitude in the angular degrees of freedom. We develop the sensor for gravitational-wave observatories, such as LIGO, Virgo, and KAGRA, to help them observe intermediate-mass black holes, increase their duty cycle, and improve localisation of sources. Our control system and its achieved sensitivity makes the sensor suitable for other fundamental physics experiments, such as tests of semiclassical gravity, searches for bosonic dark matter, and studies of the Casimir force.  
\end{abstract}

\maketitle


Ground vibrations reduce the duty cycle of interferometric gravitational-wave detectors, such as Advanced LIGO~\cite{Aasi2015} and Advanced Virgo~\cite{Acernese2014}, and limit their sensitivity below 25\,Hz~\cite{Buikema2020, Martynov_Noise_2016}. This frequency band is important for the observation of intermediate-mass black holes, accumulation of signal-to-noise ratio from lighter sources, and localisation and advanced warning of neutron star mergers for multi-messenger observations~\cite{Branchesi_2016, Yu2018}. For future detectors, such as Cosmic Explorer~\cite{Evans_CE_2021} and Einstein Telescope~\cite{Maggiore_2020}, inertial isolation is crucial for achieving the design sensitivity and for observing red-shifted signals from cosmological sources.

The LIGO test masses are suspended from actively controlled platforms, which suppress seismic vibrations in the frequency band from 100\,mHz up to 40\,Hz~\cite{Matichard_2015, MATICHARD2015273, MATICHARD2015287}. LIGO utilises commercially available Trillium 240 seismometers to measure the platform motion in its three translational and three rotational degrees of freedom. The seismometer signals are fed into six feedback loops that actuate on the platform with coil-magnet pairs. This scheme had proved successful through the LIGO detectors' observation of tens of gravitational-wave sources~\cite{LIGO_2019, Abbott2021GWTC2}. However, better inertial sensors are required to further improve the low-frequency sensitivity, duty cycle, and to simplify the lock acquisition process.




There is significant effort in the gravitational-wave community to develop new sensors, including beam rotation sensors~\cite{Venkateswara2014, Ross2020} and accelerometers with optical readout~\cite{Collette2015, Heijningen_2018}. These sensors measure platform motion along one axis and mechanically constrain the suspended mass in the other five degrees of freedom. This approach simplifies the control scheme of the platform but leads to a complex mechanical design of the system due to the need for six sensors for stabilisation.


We pursue an alternative approach with a simple mechanical design. Instead of utilising six one-axis sensors, we softly suspend a single mass without any mechanical constraints in all six degrees of freedom~\cite{MowLowry2019, Ubhi2022}. We then measure the position of the active platform relative to the suspended mass. Mechanical simplicity leads to cross-couplings between degrees of freedom; however, real-time processing can reduce the cross-couplings and enable successful stabilisation of the platform~\cite{Ubhi_2022b}. In this Letter, we discuss the design, control system, and sensitivity of the seismometer. 


\begin{figure*}
\centering
\includegraphics[width=0.95\textwidth]{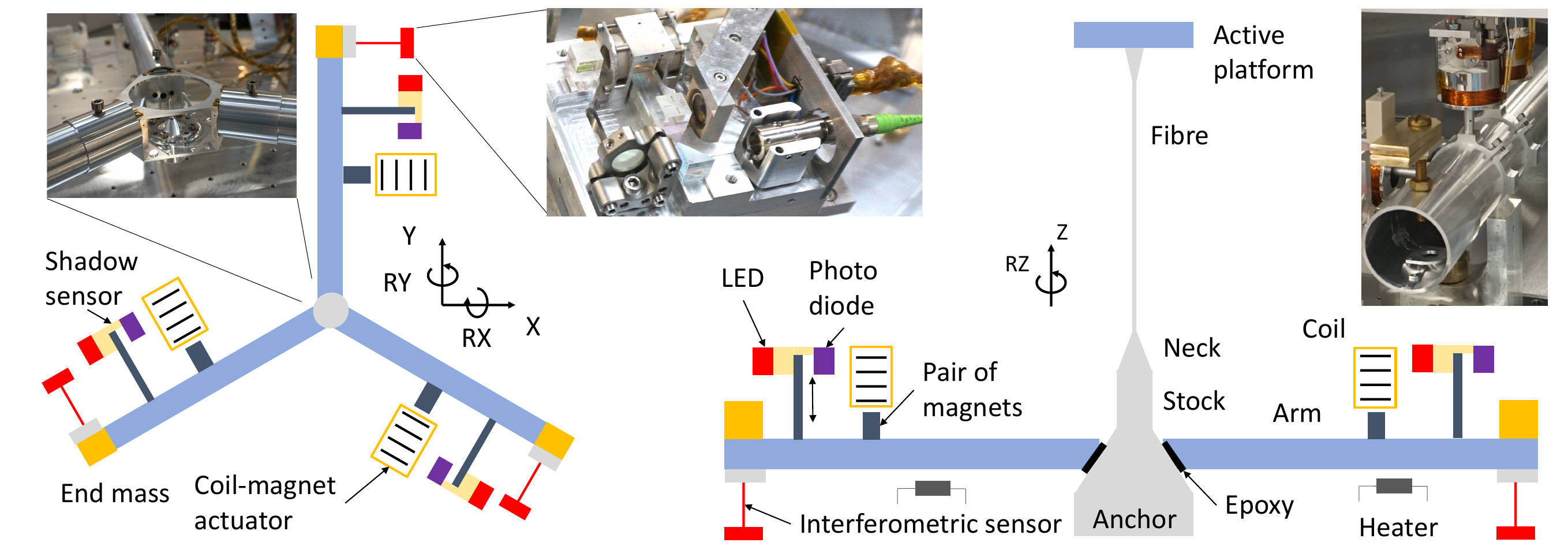}
    \caption{Scheme of the seismometer with the definition of translational (X, Y, Z) and rotational (RX, RY, RZ) axes. Left: top view (XY plane), right: side view (XZ plane). The figures show the horizontal and vertical arrangement of interferometric and shadow sensors, coil-magnet actuators and heaters.}
\label{fig:setup}
\end{figure*}

We develop the seismometer for the gravitational-wave observatories. However, the sensor may be applied in other fundamental physics experiments due to its sensitivity and stability. Particularly, in a separate experiment, we utilise the sensor and a high-finesse optical cavity to search for the signatures of semiclassical gravity~\cite{Yang_2013, Helou_2017, Liu_2023}. The sensor also has the potential to improve limits on the bosonic dark matter fields, which couple to baryon minus lepton number~\cite{Shaw_2022}. Furthermore, the seismometer is optimally set to study the Casimir forces between metamaterials~\cite{Rosa_2008, Zhao_2011}, due to its high sensitivity to torques in all rotational degrees of freedom.



\textit{Seismometer design.}- The central element of the setup is a 1-kg mass suspended with a single fused silica fibre as shown in Fig.~\ref{fig:setup}. The mass consists of three aluminium tubes with a wall thickness of 0.5\,mm and 100-g end masses made of brass. The end masses can be moved vertically to adjust the center of mass position. The mass moments of inertia around X,Y, and Z axes are $I_{zz} \approx 0.14$\,m\,kg$^2$, $I_{xx} = I_{yy} = I_{zz} / 2$. We chose non-magnetic materials for the mass to minimise the coupling of magnetic noise to the seismometer readout.

The fused silica fibre has a diameter of $80-200$\,um and a maximum stress of 2\,GPa. The fibre was pulled at the University of Glasgow with a similar procedure as the one utilised for making the LIGO fibres~\cite{Cumming2012, Cumming2020, Cumming2022}. We started with a fused silica stock (3\,mm in diameter), polished and cut it to the proper length before pulling. In the first design of the seismometer, we attached the 3-mm stock straight to the metal mass with a 1-mm thick layer of epoxy 2216 Scotchweld. We found that the layer caused a significant drift in RX and RY. This drift grew worse over time. After 6 months of operation, the mass drifted by 2\,mrad over 3 days even when the RX and RY resonances were relatively stiff (80\,mHz).

In order to reduce the drift, we welded the fibre ends to fused silica cones using a custom fused silica welding machine at the University of Glasgow. In the second design of the seismometer, we fixed the aluminium mass on the fused silica anchor with a 200\,um thick layer of indium. We chose the material to achieve a high mechanical Q-factor of the RX and RY modes because the loss angle of indium is lower compared to that of epoxy. However, we suffered from creep events~\cite{Levin_2012, Vajente_2017, Popovi_2022} in the indium layer, which was stressed at $\sim 1$\,MPa by the mass. The creep events triggered a non-stationary torque noise on the mass. As a result, the seismometer noise below 50\,mHz was modulated by temperature and fluctuated by an order of magnitude on a time scale of one day.

In the final, and successful, design of the seismometer, which is reported in this paper, we replaced indium with a 30-um thick layer of epoxy Araldite 2014-2. The creep noise has disappeared and we observed the stable and low-noise operation of the sensor. The layer of epoxy, eddy currents induced in the coil frames, and the fibre profile determined the Q-factors of the suspension modes.
We measured $Q_z = 10^4$ for the vertical (Z) mode of 7.6\,Hz, $Q_{rz} \approx 1000$ for the torsion (RZ) mode of 0.6\,mHz, set a lower limit on the Q-factors for the pendulum (X,Y) modes at 0.64\,Hz of $Q_{x,y} > 10^5$, and measured $Q_{rx,ry}=140$ for tilt modes (RX, RY) of 12--13\,mHz.


We found that the fibre rigidity is not the same for RX and RY because the fibre neck profile is elliptical in the XY plane. The asymmetry prevents the seismometer from reaching very low RX and RY modes because both of them are tuned with the same gravitational antispring~\cite{Ubhi2022}. We tuned the centre-of-mass position relative to the suspension point and achieved RX mode of 13.6\,mHz and RY mode of 12.6\,mHz. The figures imply that the difference in RX and RY fibre rigidity is $\approx 7 \times 10^{-5}$\,Nm and the minimum RX frequency we can achieve while keeping RY stable is 5\,mHz.

\textit{Sensing and control system.-} We utilised two types of sensors to measure the distance between the active platform and the suspended mass. The first is a shadow sensor~\cite{Cooper2022_BOSEM, Ubhi_2022c, Strain2012} with a linear range of 0.7\,mm and resolution of $\sim$1\,nm. The sensor consists of an LED and a photodiode mounted on the platform and a flag attached to the suspended mass as shown in Fig.~\ref{fig:setup}. Part of the LED light is blocked by the flag and its absolute position is determined by the photodiode signal.

Sensors of the second type measure the position between the mass and the platform interferometrically. We utilised custom-made Homodyne Quadrature Interferometers (HoQI) \cite{Cooper_2018, Cooper2022_L4C}. The sensors operate as Michelson interferometers with polarisation readout. By measuring the light intensity in the vertical and horizontal polarisation separately, we achieved multi-wavelength range~\cite{Cooper_2018} and better resolution compared to the shadow sensors. However, HoQIs cannot measure the absolute position of the mass relative to the platform, are more nonlinear compared to the shadow sensors, and require precise alignment ($\approx 0.2$\,mrad) of the suspended mass.

\begin{figure}
\centering
\includegraphics[width=0.45\textwidth]{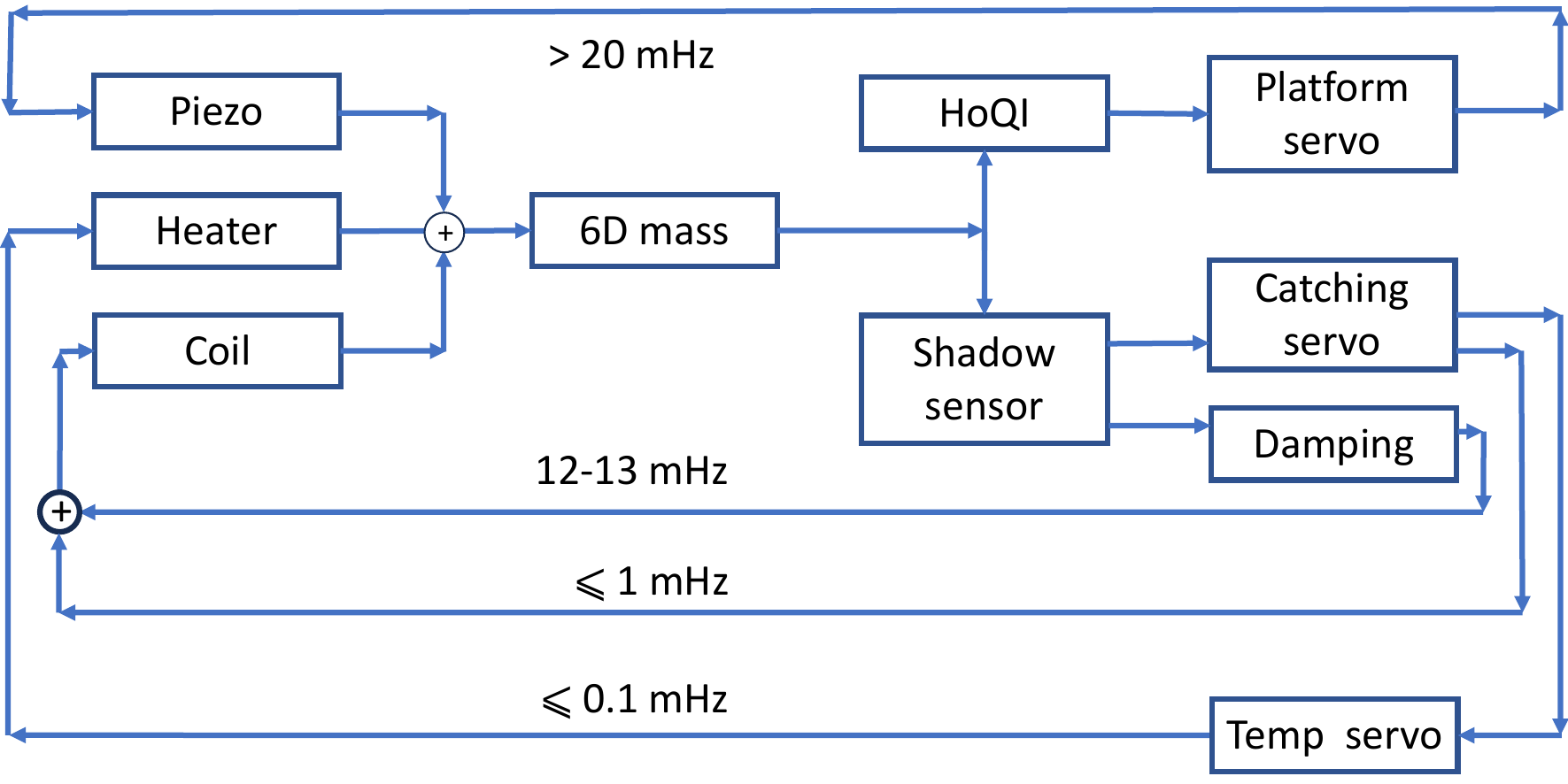}
\caption{\label{fig:feedback} Block diagram of the feedback control utilised for RX and RY. The RZ control scheme is similar but does not include the heating servo. The X,Y, and Z control schemes do not require catching loops.}
\end{figure}

In order to keep the interferometric sensors operational during temperature-induced drifts of the mass, we set up a control scheme (catching servo in Fig.~\ref{fig:feedback}) to stabilise the angular drifts below 1\,mHz. We found that if the mass is not controlled, it drifts in RX and RY by $\approx 1.5$\,mrad per 1\,K of temperature variation in the laboratory. The motion
is caused by the differential thermal expansion of the seismometer arms. The measured figures may either be explained by the imbalance of the arm lengths (by $\approx 1$\%) or by the thermal gradients along the seismometer (on the level of $\approx 40$\,mK) produced by the ambient temperature variations.


\begin{figure*}
\centering
\begin{subfigure}[b]{0.49\textwidth}
    \includegraphics[width=\textwidth]{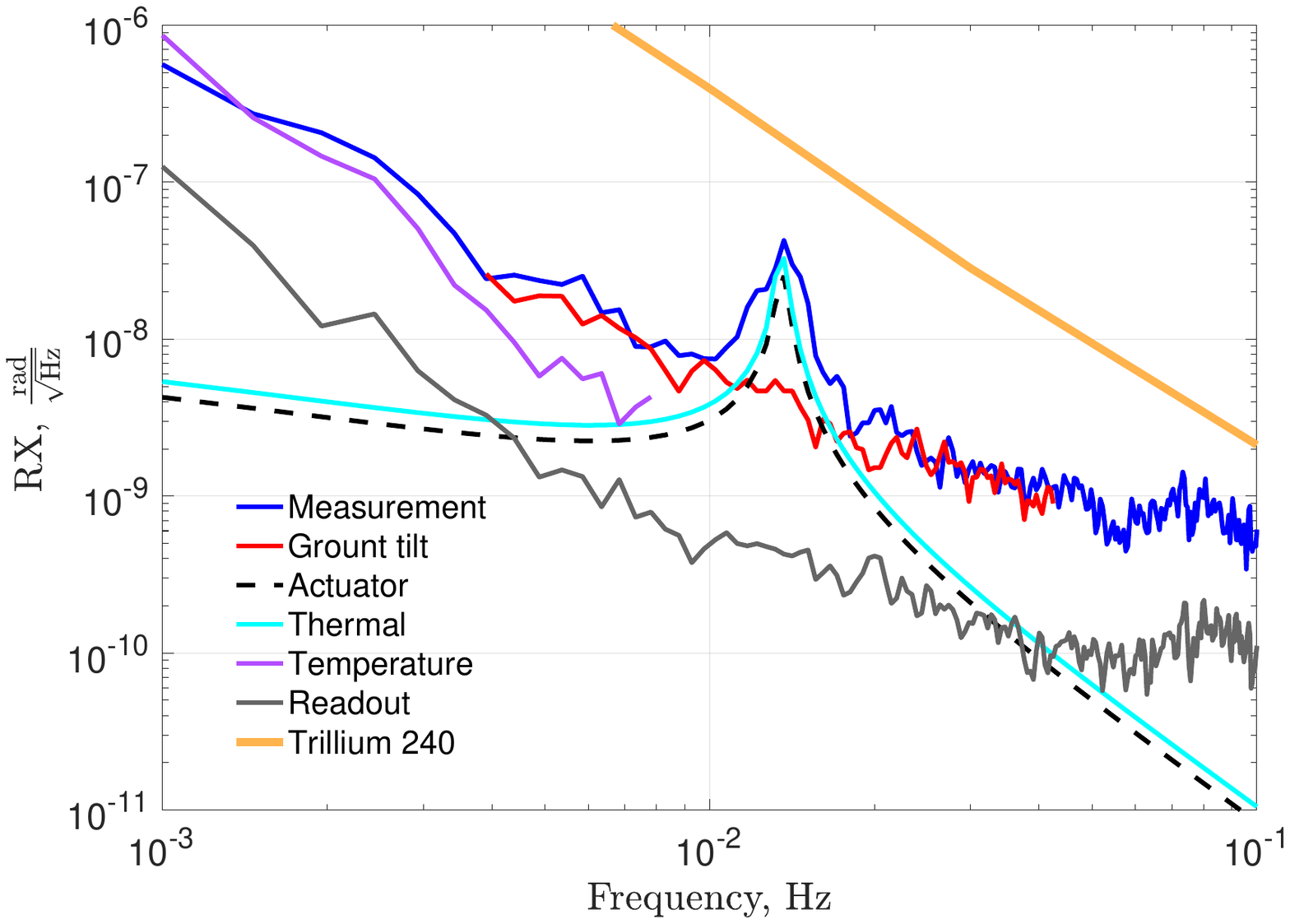}
    \caption{Tilt motion and its sensitivity analysis.}
\end{subfigure}
\begin{subfigure}[b]{0.49\textwidth}
    \includegraphics[width=\textwidth]{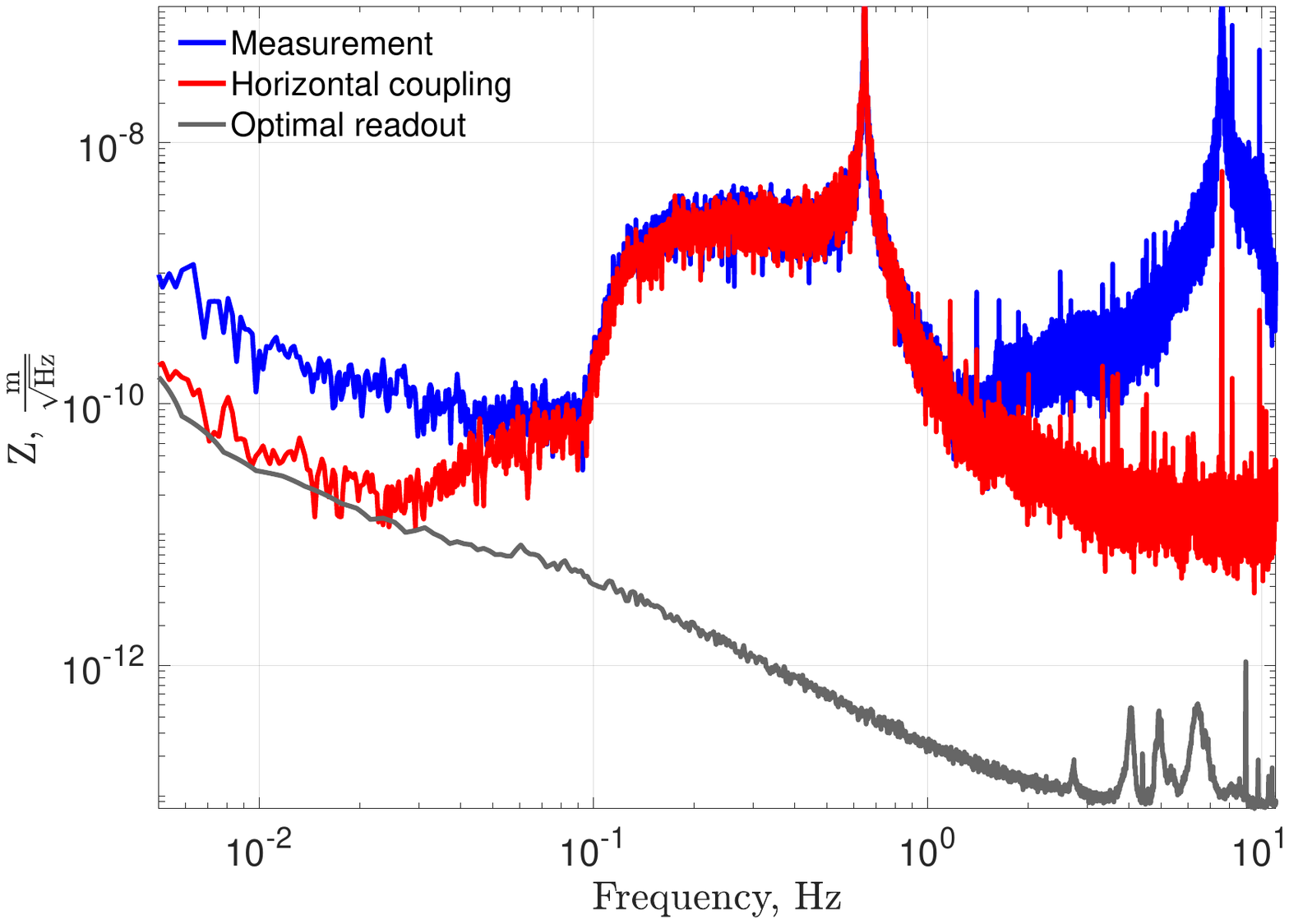}
    \caption{Vertical motion and estimation of nonlinearities.}
\end{subfigure}
\begin{subfigure}[b]{0.49\textwidth}
    \includegraphics[width=\textwidth]{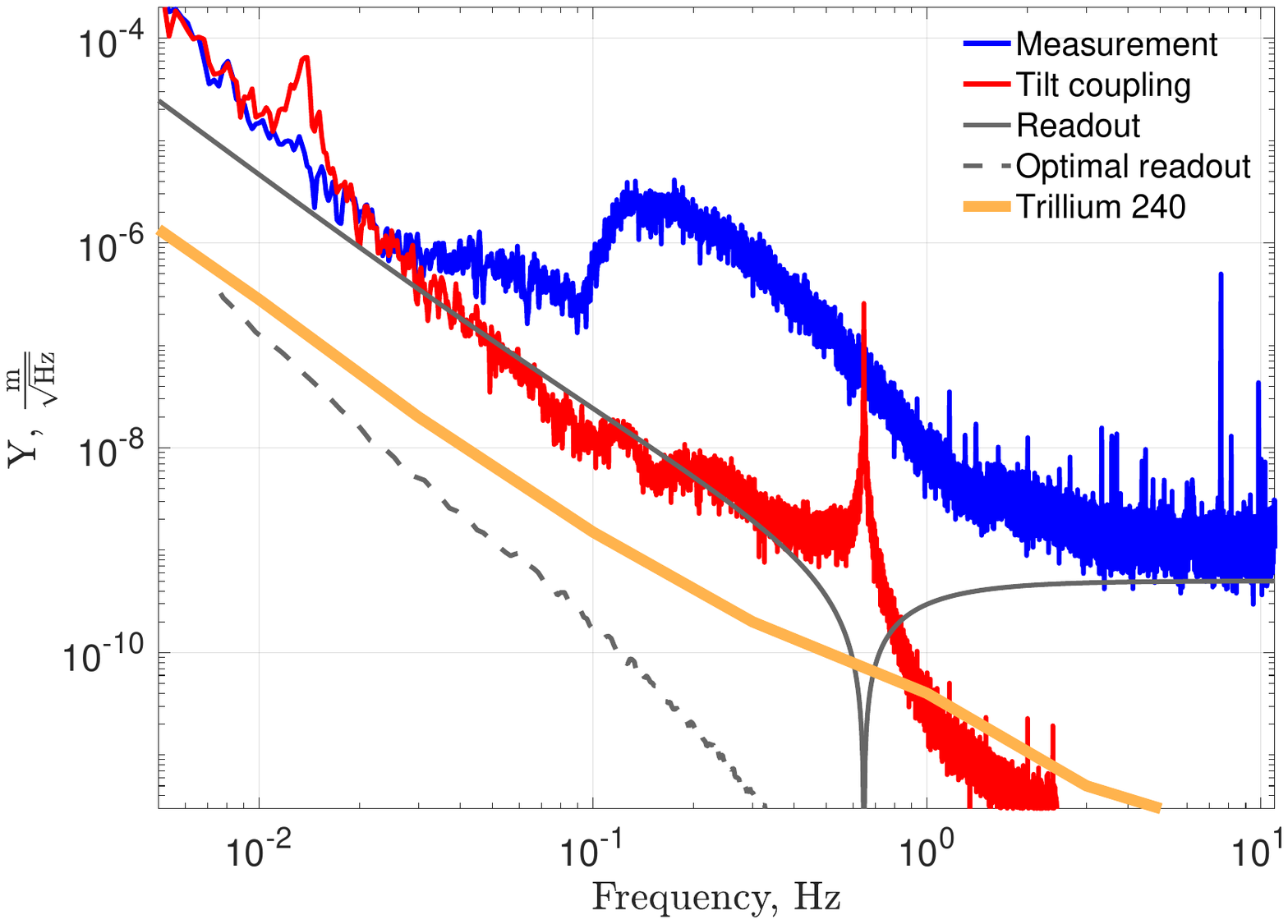}
    \caption{Horizontal motion and measured tilt coupling.}
\end{subfigure}
\begin{subfigure}[b]{0.49\textwidth}
    \includegraphics[width=\textwidth]{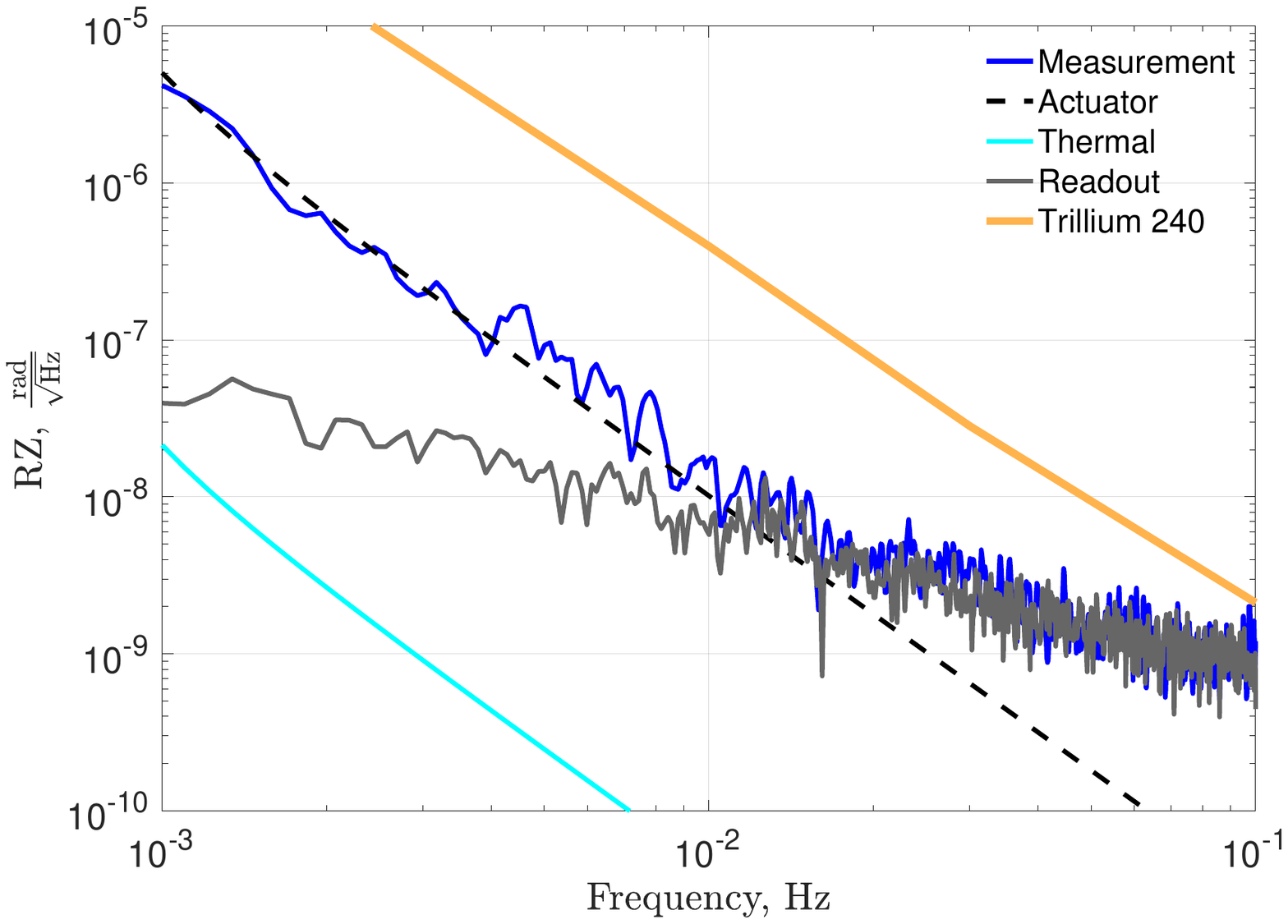}
    \caption{Torsion motion and sensitivity analysis.}
\end{subfigure}
\caption{Sensitivity of the seismometer in translational and rotational degrees of freedom. Panels a) and d) show the estimated noise of Trillium 240 seismometers separated by 1\,m to measure RX, RY, and RZ motion.}
\label{fig:asd}
\end{figure*}

Initially, we utilised only coil-magnet actuators to control the test mass position relative to the platform. For each actuation site, we mounted two 1~mm$^3$ SmCo magnets with a separation of 2\,mm. We oriented the poles of the magnets in opposite directions to minimise the coupling of ambient magnetic forces on the suspended mass. 
The coil-magnet actuation kept the shadow sensors and interferometric sensors in the linear regime. However, it also modified the dynamics of the suspended mass due to the parasitic rigidity of magnetic actuation, $k_m = dF_m / dl$, where $F_m$ is the magnetic force and $l$ is the separation between the mass and the coil. In order to achieve $k_m = 0$, the coil must be located exactly at a distance of 10\,mm from the closest magnet, and the central axes of the coil and the magnets must coincide. However, we found the condition difficult to satisfy in practice because the magnets are slightly tilted relative to each other and relative to the coil. Furthermore, the central part of the coil and its axis are poorly defined.

We have reduced the magnetic stiffness, $k_m$, by keeping the time-averaged control force $\langle F_m \rangle = 0$. We reduced the vertical coil signals to zero at frequencies below 0.1\,mHz by installing three heaters along each aluminium arm of the seismometer as shown in Fig.~\ref{fig:setup} (right). The heaters created a temperature distribution along the mass and changed the RX and RY angles by 9\,mrad per 1 Watt of heating power
provided by the 600\,Ohm resistors.  The heating scheme is non-contact and low-noise because any fluctuations of the current through the resistors are naturally filtered above 0.1\,mHz by the heat exchange process. Furthermore, the scheme does not change the dynamics of the suspended mass.

The inertial mass is stabilised relative to the active platform at frequencies below 1\,mHz. At higher frequencies, the seismometer signals can be utilised to stabilise the platform motion and to suppress ground vibrations as shown in Fig.~\ref{fig:feedback} (top path).
Conditioning of the seismometer signals before the actuation on the platform and the feedback control scheme in six axes was discussed in Ref.~\cite{Ubhi_2022b}. When the platform servo was off, we utilised the coil-magnet actuation to damp the high-Q suspension modes in translational degrees of freedom down to the Q-factors of $\sim 100$. The damping servos for the angular resonances are always on.


 
\textit{Sensitivity analysis.-}
Fig.~\ref{fig:asd} shows the spectra of the measured seismometer signals. The RX and RY sensors observe the tilt motion of the ground above 4\,mHz (as witnessed by the Y and X signals) and are limited by the thermal and actuation noise near the resonances at 12-13\,mHz. Below 4\,mHz, ambient temperature variations cause the largest noise in RX and RY. We have suppressed the noise by a factor of 5 around 1\,mHz by installing thermal shields. The shields consist of 3-mm thick aluminium panels that cover the mass from the thermal radiation of the vacuum chamber and the platform. The system passively filters ambient temperature variations at frequencies above 0.2\,mHz.

We estimated the readout noise in RX and RY sensors by measuring the vertical motion of the mass relative to the platform.
Since the Z mode is stiff, the vertical ground motion does not couple to the vertical position sensors below 1\,Hz. At 90\,mHz--1\,Hz the Z spectrum is limited by the coupling of the horizontal motion to the HoQI sensors (on the level of 1.2\%). At lower frequencies, we observe the readout noise of our vertical interferometric sensors. The noise is a factor of 5 larger than the HoQI noise measured with stationary mirrors (gray curve in Fig.~\ref{fig:asd}(b)). The additional noise comes from nonlinearities in the HoQI sensors due to the large ($\sim 0.1$\,um) motion of the platform above 5\,Hz.

The nonlinearities are caused by the ellipticity of the Lissajous figures formed by the two HoQI quadrature signals~\cite{Cooper_2018}. We define the ellipticity of the figures, $\epsilon$, as the ratio of their major and minor axes. We circularised our Lissajous figures by translating, rotating, and stretching the ellipses. However, the figures changed over time due to polarisation drifts in the optical fibres which connect the laser and the in-vacuum compact interferometers. Our long-term ellipticity was on the level of $\epsilon - 1 \approx 0.05$.
Despite noise from nonlinearities, the sensitivity of vertical interferometric sensors was still better than the one of the shadow sensors by an order of magnitude below 100\,mHz.

The noise from nonlinearities in our horizontal interferometric signals was worse than in the vertical ones by a factor of 5 due to stronger polarisation drifts in one of our in-vacuum optical fibres. The estimated readout noise in X and Y (see Fig.~\ref{fig:asd}(c)) is still low enough to measure ground vibrations, including the microseismic peak at 0.2 Hz, but is an order of magnitude worse than the Trillium 240 noise. Below 30\,mHz, the signal is limited by the tilt-to-horizontal coupling. We computed the coupling by multiplying the tilt spectrum RX by $g/\omega^2$, where $\omega$ is the angular frequency. We overestimated the tilt coupling around the RX resonance of the suspended mass because we are limited by the mass own motion rather than by the platform tilt at these frequencies as shown in Fig.~\ref{fig:asd}(a).
The RZ sensitivity (see Fig.~\ref{fig:asd}(d)) is limited by actuation noise below 10\,mHz and by readout noise at higher frequencies.



In conclusion, we have developed a six-axis seismometer that outperforms the current LIGO inertial sensors in measuring angular degrees of freedom.
Our sensitivity in the translational degrees of freedom is worse than the one of Trillium 240 due to the nonlinearities in the interferometric sensors and stiff Z mode. The nonlinearities should become smaller on the LIGO suspended platforms which move over two orders of magnitude less compared to our rigid platform above 5\,Hz. The seismometer can be further improved by (i) replacing the metal mass with a fused silica one for better thermal stability, (ii) installing a soft vertical spring to increase the seismometer response in Z, and (iii) reducing the coupling of fibre polarisation drifts to the interferometric readout by utilising a deep frequency modulation scheme~\cite{Smetana2022compact, Isleif_2019, Gerberding_2021, Gerberding_2017, Armano_2016, Weichert_2012, Rue_1972, Heinzel_2010}.

We acknowledge members of the LIGO Suspension Working Group for useful discussions, the support of the Institute for Gravitational Wave Astronomy at the University of Birmingham, STFC Equipment Call 2018 (Grant No. ST/S002154/1), STFC Consolidated Grant "Astrophysics at the University of Birmingham" (No. ST/S000305/1), and UKRI Quantum Technology for Fundamental Physics scheme (Grant No. ST/T006609/1 and ST/W006375/1). D.M. is supported by the 2021 Philip Leverhulme Prize.


\bibliography{main.bib}

\begin{thebibliography}{45}%
\makeatletter
\providecommand \@ifxundefined [1]{%
 \@ifx{#1\undefined}
}%
\providecommand \@ifnum [1]{%
 \ifnum #1\expandafter \@firstoftwo
 \else \expandafter \@secondoftwo
 \fi
}%
\providecommand \@ifx [1]{%
 \ifx #1\expandafter \@firstoftwo
 \else \expandafter \@secondoftwo
 \fi
}%
\providecommand \natexlab [1]{#1}%
\providecommand \enquote  [1]{``#1''}%
\providecommand \bibnamefont  [1]{#1}%
\providecommand \bibfnamefont [1]{#1}%
\providecommand \citenamefont [1]{#1}%
\providecommand \href@noop [0]{\@secondoftwo}%
\providecommand \href [0]{\begingroup \@sanitize@url \@href}%
\providecommand \@href[1]{\@@startlink{#1}\@@href}%
\providecommand \@@href[1]{\endgroup#1\@@endlink}%
\providecommand \@sanitize@url [0]{\catcode `\\12\catcode `\$12\catcode
  `\&12\catcode `\#12\catcode `\^12\catcode `\_12\catcode `\%12\relax}%
\providecommand \@@startlink[1]{}%
\providecommand \@@endlink[0]{}%
\providecommand \url  [0]{\begingroup\@sanitize@url \@url }%
\providecommand \@url [1]{\endgroup\@href {#1}{\urlprefix }}%
\providecommand \urlprefix  [0]{URL }%
\providecommand \Eprint [0]{\href }%
\providecommand \doibase [0]{https://doi.org/}%
\providecommand \selectlanguage [0]{\@gobble}%
\providecommand \bibinfo  [0]{\@secondoftwo}%
\providecommand \bibfield  [0]{\@secondoftwo}%
\providecommand \translation [1]{[#1]}%
\providecommand \BibitemOpen [0]{}%
\providecommand \bibitemStop [0]{}%
\providecommand \bibitemNoStop [0]{.\EOS\space}%
\providecommand \EOS [0]{\spacefactor3000\relax}%
\providecommand \BibitemShut  [1]{\csname bibitem#1\endcsname}%
\let\auto@bib@innerbib\@empty
\bibitem [{\citenamefont {Aasi}\ \emph {et~al.}(2015)\citenamefont {Aasi},
  \citenamefont {Abbott}, \citenamefont {Abbott} \emph {et~al.}}]{Aasi2015}%
  \BibitemOpen
  \bibfield  {author} {\bibinfo {author} {\bibfnamefont {J.}~\bibnamefont
  {Aasi}}, \bibinfo {author} {\bibfnamefont {B.~P.}\ \bibnamefont {Abbott}},
  \bibinfo {author} {\bibfnamefont {R.}~\bibnamefont {Abbott}}, \emph
  {et~al.},\ }\bibfield  {title} {\bibinfo {title} {Advanced ligo},\ }\href
  {https://doi.org/10.1088/0264-9381/32/7/074001} {\bibfield  {journal}
  {\bibinfo  {journal} {Classical and Quantum Gravity}\ }\textbf {\bibinfo
  {volume} {32}},\ \bibinfo {pages} {074001} (\bibinfo {year}
  {2015})}\BibitemShut {NoStop}%
\bibitem [{\citenamefont {Acernese}\ \emph {et~al.}(2014)\citenamefont
  {Acernese}, \citenamefont {Agathos}, \citenamefont {Agatsuma} \emph
  {et~al.}}]{Acernese2014}%
  \BibitemOpen
  \bibfield  {author} {\bibinfo {author} {\bibfnamefont {F.}~\bibnamefont
  {Acernese}}, \bibinfo {author} {\bibfnamefont {M.}~\bibnamefont {Agathos}},
  \bibinfo {author} {\bibfnamefont {K.}~\bibnamefont {Agatsuma}}, \emph
  {et~al.},\ }\bibfield  {title} {\bibinfo {title} {Advanced virgo: a 2nd
  generation interferometric gravitational wave detector},\ }\bibfield
  {journal} {\bibinfo  {journal} {Class. Quantum Grav. 32 (2015) 024001}\
  }\href {https://doi.org/10.1088/0264-9381/32/2/024001}
  {10.1088/0264-9381/32/2/024001} (\bibinfo {year} {2014}),\ \Eprint
  {https://arxiv.org/abs/1408.3978} {arXiv:1408.3978 [gr-qc]} \BibitemShut
  {NoStop}%
\bibitem [{\citenamefont {Buikema}\ \emph {et~al.}(2020)\citenamefont
  {Buikema}, \citenamefont {Cahillane}, \citenamefont {Mansell} \emph
  {et~al.}}]{Buikema2020}%
  \BibitemOpen
  \bibfield  {author} {\bibinfo {author} {\bibfnamefont {A.}~\bibnamefont
  {Buikema}}, \bibinfo {author} {\bibfnamefont {C.}~\bibnamefont {Cahillane}},
  \bibinfo {author} {\bibfnamefont {G.~L.}\ \bibnamefont {Mansell}}, \emph
  {et~al.} (\bibinfo {collaboration} {aLIGO}),\ }\bibfield  {title} {\bibinfo
  {title} {Sensitivity and performance of the advanced ligo detectors in the
  third observing run},\ }\href {https://doi.org/10.1103/PhysRevD.102.062003}
  {\bibfield  {journal} {\bibinfo  {journal} {Phys. Rev. D}\ }\textbf {\bibinfo
  {volume} {102}},\ \bibinfo {pages} {062003} (\bibinfo {year} {2020})},\
  \Eprint {https://arxiv.org/abs/2008.01301} {arXiv:2008.01301 [astro-ph.IM]}
  \BibitemShut {NoStop}%
\bibitem [{\citenamefont {Martynov}\ \emph {et~al.}(2016)\citenamefont
  {Martynov}, \citenamefont {Hall}, \citenamefont {Abbott} \emph
  {et~al.}}]{Martynov_Noise_2016}%
  \BibitemOpen
  \bibfield  {author} {\bibinfo {author} {\bibfnamefont {D.~V.}\ \bibnamefont
  {Martynov}}, \bibinfo {author} {\bibfnamefont {E.~D.}\ \bibnamefont {Hall}},
  \bibinfo {author} {\bibfnamefont {B.~P.}\ \bibnamefont {Abbott}}, \emph
  {et~al.},\ }\bibfield  {title} {\bibinfo {title} {Sensitivity of the advanced
  ligo detectors at the beginning of gravitational wave astronomy},\ }\href
  {https://doi.org/10.1103/PhysRevD.93.112004} {\bibfield  {journal} {\bibinfo
  {journal} {Phys. Rev. D}\ }\textbf {\bibinfo {volume} {93}},\ \bibinfo
  {pages} {112004} (\bibinfo {year} {2016})}\BibitemShut {NoStop}%
\bibitem [{\citenamefont {Branchesi}(2016)}]{Branchesi_2016}%
  \BibitemOpen
  \bibfield  {author} {\bibinfo {author} {\bibfnamefont {M.}~\bibnamefont
  {Branchesi}},\ }\bibfield  {title} {\bibinfo {title} {Multi-messenger
  astronomy: gravitational waves, neutrinos, photons, and cosmic rays},\ }\href
  {https://doi.org/10.1088/1742-6596/718/2/022004} {\bibfield  {journal}
  {\bibinfo  {journal} {Journal of Physics: Conference Series}\ }\textbf
  {\bibinfo {volume} {718}},\ \bibinfo {pages} {022004} (\bibinfo {year}
  {2016})}\BibitemShut {NoStop}%
\bibitem [{\citenamefont {Yu}\ \emph {et~al.}(2018)\citenamefont {Yu},
  \citenamefont {Martynov}, \citenamefont {Vitale} \emph {et~al.}}]{Yu2018}%
  \BibitemOpen
  \bibfield  {author} {\bibinfo {author} {\bibfnamefont {H.}~\bibnamefont
  {Yu}}, \bibinfo {author} {\bibfnamefont {D.}~\bibnamefont {Martynov}},
  \bibinfo {author} {\bibfnamefont {S.}~\bibnamefont {Vitale}}, \emph
  {et~al.},\ }\bibfield  {title} {\bibinfo {title} {Prospects for detecting
  gravitational waves at 5 hz with ground-based detectors},\ }\href
  {https://doi.org/10.1103/PhysRevLett.120.141102} {\bibfield  {journal}
  {\bibinfo  {journal} {Phys. Rev. Lett.}\ }\textbf {\bibinfo {volume} {120}},\
  \bibinfo {pages} {141102} (\bibinfo {year} {2018})}\BibitemShut {NoStop}%
\bibitem [{\citenamefont {Evans}\ \emph {et~al.}(2021)\citenamefont {Evans},
  \citenamefont {Adhikari}, \citenamefont {Afle}, \citenamefont {Ballmer},
  \citenamefont {Biscoveanu}, \citenamefont {Borhanian}, \citenamefont {Brown},
  \citenamefont {Chen}, \citenamefont {Eisenstein}, \citenamefont {Gruson},
  \citenamefont {Gupta}, \citenamefont {Hall}, \citenamefont {Huxford},
  \citenamefont {Kamai}, \citenamefont {Kashyap}, \citenamefont {Kissel},
  \citenamefont {Kuns}, \citenamefont {Landry}, \citenamefont {Lenon},
  \citenamefont {Lovelace}, \citenamefont {McCuller}, \citenamefont {Ng},
  \citenamefont {Nitz}, \citenamefont {Read}, \citenamefont {Sathyaprakash},
  \citenamefont {Shoemaker}, \citenamefont {Slagmolen}, \citenamefont {Smith},
  \citenamefont {Srivastava}, \citenamefont {Sun}, \citenamefont {Vitale},\
  and\ \citenamefont {Weiss}}]{Evans_CE_2021}%
  \BibitemOpen
  \bibfield  {author} {\bibinfo {author} {\bibfnamefont {M.}~\bibnamefont
  {Evans}}, \bibinfo {author} {\bibfnamefont {R.~X.}\ \bibnamefont {Adhikari}},
  \bibinfo {author} {\bibfnamefont {C.}~\bibnamefont {Afle}}, \bibinfo {author}
  {\bibfnamefont {S.~W.}\ \bibnamefont {Ballmer}}, \bibinfo {author}
  {\bibfnamefont {S.}~\bibnamefont {Biscoveanu}}, \bibinfo {author}
  {\bibfnamefont {S.}~\bibnamefont {Borhanian}}, \bibinfo {author}
  {\bibfnamefont {D.~A.}\ \bibnamefont {Brown}}, \bibinfo {author}
  {\bibfnamefont {Y.}~\bibnamefont {Chen}}, \bibinfo {author} {\bibfnamefont
  {R.}~\bibnamefont {Eisenstein}}, \bibinfo {author} {\bibfnamefont
  {A.}~\bibnamefont {Gruson}}, \bibinfo {author} {\bibfnamefont
  {A.}~\bibnamefont {Gupta}}, \bibinfo {author} {\bibfnamefont {E.~D.}\
  \bibnamefont {Hall}}, \bibinfo {author} {\bibfnamefont {R.}~\bibnamefont
  {Huxford}}, \bibinfo {author} {\bibfnamefont {B.}~\bibnamefont {Kamai}},
  \bibinfo {author} {\bibfnamefont {R.}~\bibnamefont {Kashyap}}, \bibinfo
  {author} {\bibfnamefont {J.~S.}\ \bibnamefont {Kissel}}, \bibinfo {author}
  {\bibfnamefont {K.}~\bibnamefont {Kuns}}, \bibinfo {author} {\bibfnamefont
  {P.}~\bibnamefont {Landry}}, \bibinfo {author} {\bibfnamefont
  {A.}~\bibnamefont {Lenon}}, \bibinfo {author} {\bibfnamefont
  {G.}~\bibnamefont {Lovelace}}, \bibinfo {author} {\bibfnamefont
  {L.}~\bibnamefont {McCuller}}, \bibinfo {author} {\bibfnamefont {K.~K.~Y.}\
  \bibnamefont {Ng}}, \bibinfo {author} {\bibfnamefont {A.~H.}\ \bibnamefont
  {Nitz}}, \bibinfo {author} {\bibfnamefont {J.}~\bibnamefont {Read}}, \bibinfo
  {author} {\bibfnamefont {B.~S.}\ \bibnamefont {Sathyaprakash}}, \bibinfo
  {author} {\bibfnamefont {D.~H.}\ \bibnamefont {Shoemaker}}, \bibinfo {author}
  {\bibfnamefont {B.~J.~J.}\ \bibnamefont {Slagmolen}}, \bibinfo {author}
  {\bibfnamefont {J.~R.}\ \bibnamefont {Smith}}, \bibinfo {author}
  {\bibfnamefont {V.}~\bibnamefont {Srivastava}}, \bibinfo {author}
  {\bibfnamefont {L.}~\bibnamefont {Sun}}, \bibinfo {author} {\bibfnamefont
  {S.}~\bibnamefont {Vitale}},\ and\ \bibinfo {author} {\bibfnamefont
  {R.}~\bibnamefont {Weiss}},\ }\href@noop {} {\bibinfo {title} {A horizon
  study for cosmic explorer: Science, observatories, and community}} (\bibinfo
  {year} {2021}),\ \Eprint {https://arxiv.org/abs/2109.09882} {arXiv:2109.09882
  [astro-ph.IM]} \BibitemShut {NoStop}%
\bibitem [{\citenamefont {Maggiore}\ \emph {et~al.}(2020)\citenamefont
  {Maggiore}, \citenamefont {Broeck}, \citenamefont {Bartolo}, \citenamefont
  {Belgacem}, \citenamefont {Bertacca}, \citenamefont {Bizouard}, \citenamefont
  {Branchesi}, \citenamefont {Clesse}, \citenamefont {Foffa}, \citenamefont
  {Garc{\'{\i} }a-Bellido}, \citenamefont {Grimm}, \citenamefont {Harms},
  \citenamefont {Hinderer}, \citenamefont {Matarrese}, \citenamefont {Palomba},
  \citenamefont {Peloso}, \citenamefont {Ricciardone},\ and\ \citenamefont
  {Sakellariadou}}]{Maggiore_2020}%
  \BibitemOpen
  \bibfield  {author} {\bibinfo {author} {\bibfnamefont {M.}~\bibnamefont
  {Maggiore}}, \bibinfo {author} {\bibfnamefont {C.~V.~D.}\ \bibnamefont
  {Broeck}}, \bibinfo {author} {\bibfnamefont {N.}~\bibnamefont {Bartolo}},
  \bibinfo {author} {\bibfnamefont {E.}~\bibnamefont {Belgacem}}, \bibinfo
  {author} {\bibfnamefont {D.}~\bibnamefont {Bertacca}}, \bibinfo {author}
  {\bibfnamefont {M.~A.}\ \bibnamefont {Bizouard}}, \bibinfo {author}
  {\bibfnamefont {M.}~\bibnamefont {Branchesi}}, \bibinfo {author}
  {\bibfnamefont {S.}~\bibnamefont {Clesse}}, \bibinfo {author} {\bibfnamefont
  {S.}~\bibnamefont {Foffa}}, \bibinfo {author} {\bibfnamefont
  {J.}~\bibnamefont {Garc{\'{\i} }a-Bellido}}, \bibinfo {author} {\bibfnamefont
  {S.}~\bibnamefont {Grimm}}, \bibinfo {author} {\bibfnamefont
  {J.}~\bibnamefont {Harms}}, \bibinfo {author} {\bibfnamefont
  {T.}~\bibnamefont {Hinderer}}, \bibinfo {author} {\bibfnamefont
  {S.}~\bibnamefont {Matarrese}}, \bibinfo {author} {\bibfnamefont
  {C.}~\bibnamefont {Palomba}}, \bibinfo {author} {\bibfnamefont
  {M.}~\bibnamefont {Peloso}}, \bibinfo {author} {\bibfnamefont
  {A.}~\bibnamefont {Ricciardone}},\ and\ \bibinfo {author} {\bibfnamefont
  {M.}~\bibnamefont {Sakellariadou}},\ }\bibfield  {title} {\bibinfo {title}
  {Science case for the einstein telescope},\ }\href
  {https://doi.org/10.1088/1475-7516/2020/03/050} {\bibfield  {journal}
  {\bibinfo  {journal} {Journal of Cosmology and Astroparticle Physics}\
  }\textbf {\bibinfo {volume} {2020}}\bibinfo  {number} { (03)},\ \bibinfo
  {pages} {050}}\BibitemShut {NoStop}%
\bibitem [{\citenamefont {Matichard}\ \emph
  {et~al.}(2015{\natexlab{a}})\citenamefont {Matichard}, \citenamefont {Lantz},
  \citenamefont {Mittleman} \emph {et~al.}}]{Matichard_2015}%
  \BibitemOpen
\bibfield  {number} {  }\bibfield  {author} {\bibinfo {author} {\bibfnamefont
  {F.}~\bibnamefont {Matichard}}, \bibinfo {author} {\bibfnamefont
  {B.}~\bibnamefont {Lantz}}, \bibinfo {author} {\bibfnamefont
  {R.}~\bibnamefont {Mittleman}}, \emph {et~al.},\ }\bibfield  {title}
  {\bibinfo {title} {Seismic isolation of advanced {LIGO}: Review of strategy,
  instrumentation and performance},\ }\href
  {https://doi.org/10.1088/0264-9381/32/18/185003} {\bibfield  {journal}
  {\bibinfo  {journal} {Classical and Quantum Gravity}\ }\textbf {\bibinfo
  {volume} {32}},\ \bibinfo {pages} {185003} (\bibinfo {year}
  {2015}{\natexlab{a}})}\BibitemShut {NoStop}%
\bibitem [{\citenamefont {Matichard}\ \emph
  {et~al.}(2015{\natexlab{b}})\citenamefont {Matichard}, \citenamefont {Lantz},
  \citenamefont {Mason} \emph {et~al.}}]{MATICHARD2015273}%
  \BibitemOpen
  \bibfield  {author} {\bibinfo {author} {\bibfnamefont {F.}~\bibnamefont
  {Matichard}}, \bibinfo {author} {\bibfnamefont {B.}~\bibnamefont {Lantz}},
  \bibinfo {author} {\bibfnamefont {K.}~\bibnamefont {Mason}}, \emph {et~al.},\
  }\bibfield  {title} {\bibinfo {title} {Advanced ligo two-stage twelve-axis
  vibration isolation and positioning platform. part 1: Design and production
  overview},\ }\href
  {https://doi.org/https://doi.org/10.1016/j.precisioneng.2014.09.010}
  {\bibfield  {journal} {\bibinfo  {journal} {Precision Engineering}\ }\textbf
  {\bibinfo {volume} {40}},\ \bibinfo {pages} {273} (\bibinfo {year}
  {2015}{\natexlab{b}})}\BibitemShut {NoStop}%
\bibitem [{\citenamefont {Matichard}\ \emph
  {et~al.}(2015{\natexlab{c}})\citenamefont {Matichard}, \citenamefont {Lantz},
  \citenamefont {Mason} \emph {et~al.}}]{MATICHARD2015287}%
  \BibitemOpen
  \bibfield  {author} {\bibinfo {author} {\bibfnamefont {F.}~\bibnamefont
  {Matichard}}, \bibinfo {author} {\bibfnamefont {B.}~\bibnamefont {Lantz}},
  \bibinfo {author} {\bibfnamefont {K.}~\bibnamefont {Mason}}, \emph {et~al.},\
  }\bibfield  {title} {\bibinfo {title} {Advanced ligo two-stage twelve-axis
  vibration isolation and positioning platform. part 2: Experimental
  investigation and tests results},\ }\href
  {https://doi.org/https://doi.org/10.1016/j.precisioneng.2014.11.010}
  {\bibfield  {journal} {\bibinfo  {journal} {Precision Engineering}\ }\textbf
  {\bibinfo {volume} {40}},\ \bibinfo {pages} {287 } (\bibinfo {year}
  {2015}{\natexlab{c}})}\BibitemShut {NoStop}%
\bibitem [{\citenamefont {Collaboration}\ and\ \citenamefont
  {Collaboration}(2019)}]{LIGO_2019}%
  \BibitemOpen
  \bibfield  {author} {\bibinfo {author} {\bibfnamefont {L.~S.}\ \bibnamefont
  {Collaboration}}\ and\ \bibinfo {author} {\bibfnamefont {V.}~\bibnamefont
  {Collaboration}},\ }\bibfield  {title} {\bibinfo {title} {Gwtc-1: A
  gravitational-wave transient catalog of compact binary mergers observed by
  ligo and virgo during the first and second observing runs},\ }\href
  {https://doi.org/10.1103/PhysRevX.9.031040} {\bibfield  {journal} {\bibinfo
  {journal} {Phys. Rev. X}\ }\textbf {\bibinfo {volume} {9}},\ \bibinfo {pages}
  {031040} (\bibinfo {year} {2019})}\BibitemShut {NoStop}%
\bibitem [{\citenamefont {Abbott}\ \emph {et~al.}(2021)\citenamefont {Abbott},
  \citenamefont {Abbott}, \citenamefont {Abraham} \emph
  {et~al.}}]{Abbott2021GWTC2}%
  \BibitemOpen
  \bibfield  {author} {\bibinfo {author} {\bibfnamefont {R.}~\bibnamefont
  {Abbott}}, \bibinfo {author} {\bibfnamefont {T.~D.}\ \bibnamefont {Abbott}},
  \bibinfo {author} {\bibfnamefont {S.}~\bibnamefont {Abraham}}, \emph {et~al.}
  (\bibinfo {collaboration} {LIGO Scientific Collaboration and Virgo
  Collaboration}),\ }\bibfield  {title} {\bibinfo {title} {Gwtc-2: Compact
  binary coalescences observed by ligo and virgo during the first half of the
  third observing run},\ }\href {https://doi.org/10.1103/PhysRevX.11.021053}
  {\bibfield  {journal} {\bibinfo  {journal} {Phys. Rev. X}\ }\textbf {\bibinfo
  {volume} {11}},\ \bibinfo {pages} {021053} (\bibinfo {year}
  {2021})}\BibitemShut {NoStop}%
\bibitem [{\citenamefont {Venkateswara}\ \emph {et~al.}(2014)\citenamefont
  {Venkateswara}, \citenamefont {Hagedorn}, \citenamefont {Turner} \emph
  {et~al.}}]{Venkateswara2014}%
  \BibitemOpen
  \bibfield  {author} {\bibinfo {author} {\bibfnamefont {K.}~\bibnamefont
  {Venkateswara}}, \bibinfo {author} {\bibfnamefont {C.}~\bibnamefont
  {Hagedorn}}, \bibinfo {author} {\bibfnamefont {M.}~\bibnamefont {Turner}},
  \emph {et~al.},\ }\bibfield  {title} {\bibinfo {title} {A high-precision
  mechanical absolute-rotation sensor},\ }\href
  {https://doi.org/10.1063/1.4862816} {\bibfield  {journal} {\bibinfo
  {journal} {The Review of scientific instruments}\ }\textbf {\bibinfo {volume}
  {85}},\ \bibinfo {pages} {015005} (\bibinfo {year} {2014})}\BibitemShut
  {NoStop}%
\bibitem [{\citenamefont {Ross}\ \emph {et~al.}(2020)\citenamefont {Ross},
  \citenamefont {Venkateswara}, \citenamefont {Mow-Lowry} \emph
  {et~al.}}]{Ross2020}%
  \BibitemOpen
  \bibfield  {author} {\bibinfo {author} {\bibfnamefont {M.~P.}\ \bibnamefont
  {Ross}}, \bibinfo {author} {\bibfnamefont {K.}~\bibnamefont {Venkateswara}},
  \bibinfo {author} {\bibfnamefont {C.}~\bibnamefont {Mow-Lowry}}, \emph
  {et~al.},\ }\bibfield  {title} {\bibinfo {title} {Towards windproofing
  {LIGO}: reducing the effect of wind-driven floor tilt by using rotation
  sensors in active seismic isolation},\ }\href
  {https://doi.org/10.1088/1361-6382/ab9d5c} {\bibfield  {journal} {\bibinfo
  {journal} {Classical and Quantum Gravity}\ }\textbf {\bibinfo {volume}
  {37}},\ \bibinfo {pages} {185018} (\bibinfo {year} {2020})}\BibitemShut
  {NoStop}%
\bibitem [{\citenamefont {Collette}\ \emph {et~al.}(2015)\citenamefont
  {Collette}, \citenamefont {Nassif}, \citenamefont {Amar} \emph
  {et~al.}}]{Collette2015}%
  \BibitemOpen
  \bibfield  {author} {\bibinfo {author} {\bibfnamefont {C.}~\bibnamefont
  {Collette}}, \bibinfo {author} {\bibfnamefont {F.}~\bibnamefont {Nassif}},
  \bibinfo {author} {\bibfnamefont {J.}~\bibnamefont {Amar}}, \emph {et~al.},\
  }\bibfield  {title} {\bibinfo {title} {Prototype of interferometric absolute
  motion sensor},\ }\href@noop {} {\bibfield  {journal} {\bibinfo  {journal}
  {Sensors and Actuators, A: Physical}\ }\textbf {\bibinfo {volume} {224}},\
  \bibinfo {pages} {72} (\bibinfo {year} {2015})}\BibitemShut {NoStop}%
\bibitem [{\citenamefont {van Heijningen}\ \emph {et~al.}(2018)\citenamefont
  {van Heijningen}, \citenamefont {Bertolini},\ and\ \citenamefont {van~den
  Brand}}]{Heijningen_2018}%
  \BibitemOpen
  \bibfield  {author} {\bibinfo {author} {\bibfnamefont {J.~V.}\ \bibnamefont
  {van Heijningen}}, \bibinfo {author} {\bibfnamefont {A.}~\bibnamefont
  {Bertolini}},\ and\ \bibinfo {author} {\bibfnamefont {J.~F.~J.}\ \bibnamefont
  {van~den Brand}},\ }\bibfield  {title} {\bibinfo {title} {A novel
  interferometrically read out inertial sensor for future gravitational wave
  detectors},\ }in\ \href {https://doi.org/10.1109/SAS.2018.8336722} {\emph
  {\bibinfo {booktitle} {2018 IEEE Sensors Applications Symposium (SAS)}}}\
  (\bibinfo {year} {2018})\ pp.\ \bibinfo {pages} {1--5}\BibitemShut {NoStop}%
\bibitem [{\citenamefont {Mow-Lowry}\ and\ \citenamefont
  {Martynov}(2019)}]{MowLowry2019}%
  \BibitemOpen
  \bibfield  {author} {\bibinfo {author} {\bibfnamefont {C.~M.}\ \bibnamefont
  {Mow-Lowry}}\ and\ \bibinfo {author} {\bibfnamefont {D.}~\bibnamefont
  {Martynov}},\ }\bibfield  {title} {\bibinfo {title} {A 6d interferometric
  inertial isolation system},\ }\href
  {https://doi.org/10.1088/1361-6382/ab4e01} {\bibfield  {journal} {\bibinfo
  {journal} {Classical and Quantum Gravity}\ }\textbf {\bibinfo {volume}
  {36}},\ \bibinfo {eid} {245006} (\bibinfo {year} {2019})},\ \Eprint
  {https://arxiv.org/abs/1801.01468} {arXiv:1801.01468 [astro-ph.IM]}
  \BibitemShut {NoStop}%
\bibitem [{\citenamefont {Ubhi}\ \emph
  {et~al.}(2022{\natexlab{a}})\citenamefont {Ubhi}, \citenamefont {Smetana},
  \citenamefont {Zhang} \emph {et~al.}}]{Ubhi2022}%
  \BibitemOpen
  \bibfield  {author} {\bibinfo {author} {\bibfnamefont {A.~S.}\ \bibnamefont
  {Ubhi}}, \bibinfo {author} {\bibfnamefont {J.}~\bibnamefont {Smetana}},
  \bibinfo {author} {\bibfnamefont {T.}~\bibnamefont {Zhang}}, \emph {et~al.},\
  }\bibfield  {title} {\bibinfo {title} {A six degree-of-freedom fused silica
  seismometer: design and tests of a metal prototype},\ }\href
  {https://doi.org/10.1088/1361-6382/ac39b9} {\bibfield  {journal} {\bibinfo
  {journal} {Class. Quant. Grav.}\ }\textbf {\bibinfo {volume} {39}},\ \bibinfo
  {pages} {015006} (\bibinfo {year} {2022}{\natexlab{a}})},\ \Eprint
  {https://arxiv.org/abs/2109.07880} {arXiv:2109.07880 [physics.ins-det]}
  \BibitemShut {NoStop}%
\bibitem [{\citenamefont {Ubhi}\ \emph
  {et~al.}(2022{\natexlab{b}})\citenamefont {Ubhi}, \citenamefont {Prokhorov},
  \citenamefont {Cooper}, \citenamefont {Fronzo}, \citenamefont {Bryant},
  \citenamefont {Hoyland}, \citenamefont {Mitchell}, \citenamefont
  {Van~Dongen}, \citenamefont {Mow-Lowry}, \citenamefont {Cumming} \emph
  {et~al.}}]{Ubhi_2022b}%
  \BibitemOpen
  \bibfield  {author} {\bibinfo {author} {\bibfnamefont {A.~S.}\ \bibnamefont
  {Ubhi}}, \bibinfo {author} {\bibfnamefont {L.}~\bibnamefont {Prokhorov}},
  \bibinfo {author} {\bibfnamefont {S.}~\bibnamefont {Cooper}}, \bibinfo
  {author} {\bibfnamefont {C.~D.}\ \bibnamefont {Fronzo}}, \bibinfo {author}
  {\bibfnamefont {J.}~\bibnamefont {Bryant}}, \bibinfo {author} {\bibfnamefont
  {D.}~\bibnamefont {Hoyland}}, \bibinfo {author} {\bibfnamefont
  {A.}~\bibnamefont {Mitchell}}, \bibinfo {author} {\bibfnamefont
  {J.}~\bibnamefont {Van~Dongen}}, \bibinfo {author} {\bibfnamefont
  {C.}~\bibnamefont {Mow-Lowry}}, \bibinfo {author} {\bibfnamefont
  {A.}~\bibnamefont {Cumming}}, \emph {et~al.},\ }\bibfield  {title} {\bibinfo
  {title} {Active platform stabilization with a 6d seismometer},\ }\href@noop
  {} {\bibfield  {journal} {\bibinfo  {journal} {Applied Physics Letters}\
  }\textbf {\bibinfo {volume} {121}},\ \bibinfo {pages} {174101} (\bibinfo
  {year} {2022}{\natexlab{b}})}\BibitemShut {NoStop}%
\bibitem [{\citenamefont {Yang}\ \emph {et~al.}(2013)\citenamefont {Yang},
  \citenamefont {Miao}, \citenamefont {Lee}, \citenamefont {Helou},\ and\
  \citenamefont {Chen}}]{Yang_2013}%
  \BibitemOpen
  \bibfield  {author} {\bibinfo {author} {\bibfnamefont {H.}~\bibnamefont
  {Yang}}, \bibinfo {author} {\bibfnamefont {H.}~\bibnamefont {Miao}}, \bibinfo
  {author} {\bibfnamefont {D.-S.}\ \bibnamefont {Lee}}, \bibinfo {author}
  {\bibfnamefont {B.}~\bibnamefont {Helou}},\ and\ \bibinfo {author}
  {\bibfnamefont {Y.}~\bibnamefont {Chen}},\ }\bibfield  {title} {\bibinfo
  {title} {Macroscopic quantum mechanics in a classical spacetime},\ }\bibfield
   {journal} {\bibinfo  {journal} {Physical Review Letters}\ }\textbf {\bibinfo
  {volume} {110}},\ \href {https://doi.org/10.1103/physrevlett.110.170401}
  {10.1103/physrevlett.110.170401} (\bibinfo {year} {2013})\BibitemShut
  {NoStop}%
\bibitem [{\citenamefont {Helou}\ \emph {et~al.}(2017)\citenamefont {Helou},
  \citenamefont {Luo}, \citenamefont {Yeh}, \citenamefont {gang Shao},
  \citenamefont {Slagmolen}, \citenamefont {McClelland},\ and\ \citenamefont
  {Chen}}]{Helou_2017}%
  \BibitemOpen
  \bibfield  {author} {\bibinfo {author} {\bibfnamefont {B.}~\bibnamefont
  {Helou}}, \bibinfo {author} {\bibfnamefont {J.}~\bibnamefont {Luo}}, \bibinfo
  {author} {\bibfnamefont {H.-C.}\ \bibnamefont {Yeh}}, \bibinfo {author}
  {\bibfnamefont {C.}~\bibnamefont {gang Shao}}, \bibinfo {author}
  {\bibfnamefont {B.}~\bibnamefont {Slagmolen}}, \bibinfo {author}
  {\bibfnamefont {D.~E.}\ \bibnamefont {McClelland}},\ and\ \bibinfo {author}
  {\bibfnamefont {Y.}~\bibnamefont {Chen}},\ }\bibfield  {title} {\bibinfo
  {title} {Measurable signatures of quantum mechanics in a classical
  spacetime},\ }\bibfield  {journal} {\bibinfo  {journal} {Physical Review D}\
  }\textbf {\bibinfo {volume} {96}},\ \href
  {https://doi.org/10.1103/physrevd.96.044008} {10.1103/physrevd.96.044008}
  (\bibinfo {year} {2017})\BibitemShut {NoStop}%
\bibitem [{\citenamefont {Liu}\ \emph {et~al.}(2023)\citenamefont {Liu},
  \citenamefont {Miao}, \citenamefont {Chen},\ and\ \citenamefont
  {Ma}}]{Liu_2023}%
  \BibitemOpen
  \bibfield  {author} {\bibinfo {author} {\bibfnamefont {Y.}~\bibnamefont
  {Liu}}, \bibinfo {author} {\bibfnamefont {H.}~\bibnamefont {Miao}}, \bibinfo
  {author} {\bibfnamefont {Y.}~\bibnamefont {Chen}},\ and\ \bibinfo {author}
  {\bibfnamefont {Y.}~\bibnamefont {Ma}},\ }\bibfield  {title} {\bibinfo
  {title} {Semiclassical gravity phenomenology under the causal-conditional
  quantum measurement prescription},\ }\bibfield  {journal} {\bibinfo
  {journal} {Physical Review D}\ }\textbf {\bibinfo {volume} {107}},\ \href
  {https://doi.org/10.1103/physrevd.107.024004} {10.1103/physrevd.107.024004}
  (\bibinfo {year} {2023})\BibitemShut {NoStop}%
\bibitem [{\citenamefont {Shaw}\ \emph {et~al.}(2022)\citenamefont {Shaw},
  \citenamefont {Ross}, \citenamefont {Hagedorn}, \citenamefont {Adelberger},\
  and\ \citenamefont {Gundlach}}]{Shaw_2022}%
  \BibitemOpen
  \bibfield  {author} {\bibinfo {author} {\bibfnamefont {E.}~\bibnamefont
  {Shaw}}, \bibinfo {author} {\bibfnamefont {M.}~\bibnamefont {Ross}}, \bibinfo
  {author} {\bibfnamefont {C.}~\bibnamefont {Hagedorn}}, \bibinfo {author}
  {\bibfnamefont {E.}~\bibnamefont {Adelberger}},\ and\ \bibinfo {author}
  {\bibfnamefont {J.}~\bibnamefont {Gundlach}},\ }\bibfield  {title} {\bibinfo
  {title} {Torsion-balance search for ultralow-mass bosonic dark matter},\
  }\bibfield  {journal} {\bibinfo  {journal} {Physical Review D}\ }\textbf
  {\bibinfo {volume} {105}},\ \href
  {https://doi.org/10.1103/physrevd.105.042007} {10.1103/physrevd.105.042007}
  (\bibinfo {year} {2022})\BibitemShut {NoStop}%
\bibitem [{\citenamefont {Rosa}\ \emph {et~al.}(2008)\citenamefont {Rosa},
  \citenamefont {Dalvit},\ and\ \citenamefont {Milonni}}]{Rosa_2008}%
  \BibitemOpen
  \bibfield  {author} {\bibinfo {author} {\bibfnamefont {F.~S.~S.}\
  \bibnamefont {Rosa}}, \bibinfo {author} {\bibfnamefont {D.~A.~R.}\
  \bibnamefont {Dalvit}},\ and\ \bibinfo {author} {\bibfnamefont {P.~W.}\
  \bibnamefont {Milonni}},\ }\bibfield  {title} {\bibinfo {title}
  {Casimir-lifshitz theory and metamaterials},\ }\bibfield  {journal} {\bibinfo
   {journal} {Physical Review Letters}\ }\textbf {\bibinfo {volume} {100}},\
  \href {https://doi.org/10.1103/physrevlett.100.183602}
  {10.1103/physrevlett.100.183602} (\bibinfo {year} {2008})\BibitemShut
  {NoStop}%
\bibitem [{\citenamefont {Zhao}\ and\ \citenamefont {Miao}(2011)}]{Zhao_2011}%
  \BibitemOpen
  \bibfield  {author} {\bibinfo {author} {\bibfnamefont {T.-M.}\ \bibnamefont
  {Zhao}}\ and\ \bibinfo {author} {\bibfnamefont {R.-X.}\ \bibnamefont
  {Miao}},\ }\bibfield  {title} {\bibinfo {title} {Huge casimir effect at
  finite temperature in electromagnetic rindler space},\ }\href
  {https://doi.org/10.1364/ol.36.004467} {\bibfield  {journal} {\bibinfo
  {journal} {Optics Letters}\ }\textbf {\bibinfo {volume} {36}},\ \bibinfo
  {pages} {4467} (\bibinfo {year} {2011})}\BibitemShut {NoStop}%
\bibitem [{\citenamefont {Cumming}\ \emph {et~al.}(2012)\citenamefont
  {Cumming}, \citenamefont {Bell}, \citenamefont {Barsotti} \emph
  {et~al.}}]{Cumming2012}%
  \BibitemOpen
  \bibfield  {author} {\bibinfo {author} {\bibfnamefont {A.~V.}\ \bibnamefont
  {Cumming}}, \bibinfo {author} {\bibfnamefont {A.~S.}\ \bibnamefont {Bell}},
  \bibinfo {author} {\bibfnamefont {L.}~\bibnamefont {Barsotti}}, \emph
  {et~al.},\ }\bibfield  {title} {\bibinfo {title} {Design and development of
  the advanced {LIGO} monolithic fused silica suspension},\ }\href
  {https://doi.org/10.1088/0264-9381/29/3/035003} {\bibfield  {journal}
  {\bibinfo  {journal} {Classical and Quantum Gravity}\ }\textbf {\bibinfo
  {volume} {29}},\ \bibinfo {pages} {035003} (\bibinfo {year}
  {2012})}\BibitemShut {NoStop}%
\bibitem [{\citenamefont {Cumming}\ \emph {et~al.}(2020)\citenamefont
  {Cumming}, \citenamefont {Sorazu}, \citenamefont {Daw} \emph
  {et~al.}}]{Cumming2020}%
  \BibitemOpen
  \bibfield  {author} {\bibinfo {author} {\bibfnamefont {A.~V.}\ \bibnamefont
  {Cumming}}, \bibinfo {author} {\bibfnamefont {B.}~\bibnamefont {Sorazu}},
  \bibinfo {author} {\bibfnamefont {E.}~\bibnamefont {Daw}}, \emph {et~al.},\
  }\bibfield  {title} {\bibinfo {title} {Lowest observed surface and weld
  losses in fused silica fibres for gravitational wave detectors},\ }\href
  {https://doi.org/10.1088/1361-6382/abac42} {\bibfield  {journal} {\bibinfo
  {journal} {Class. Quant. Grav.}\ }\textbf {\bibinfo {volume} {37}},\ \bibinfo
  {pages} {195019} (\bibinfo {year} {2020})}\BibitemShut {NoStop}%
\bibitem [{\citenamefont {Cumming}\ \emph {et~al.}(2022)\citenamefont
  {Cumming}, \citenamefont {Jones}, \citenamefont {Hammond} \emph
  {et~al.}}]{Cumming2022}%
  \BibitemOpen
  \bibfield  {author} {\bibinfo {author} {\bibfnamefont {A.~V.}\ \bibnamefont
  {Cumming}}, \bibinfo {author} {\bibfnamefont {R.}~\bibnamefont {Jones}},
  \bibinfo {author} {\bibfnamefont {G.~D.}\ \bibnamefont {Hammond}}, \emph
  {et~al.},\ }\bibfield  {title} {\bibinfo {title} {Large-scale monolithic
  fused-silica mirror suspension for third-generation gravitational-wave
  detectors},\ }\href {https://doi.org/10.1103/PhysRevApplied.17.024044}
  {\bibfield  {journal} {\bibinfo  {journal} {Phys. Rev. Applied}\ }\textbf
  {\bibinfo {volume} {17}},\ \bibinfo {pages} {024044} (\bibinfo {year}
  {2022})}\BibitemShut {NoStop}%
\bibitem [{\citenamefont {Levin}(2012)}]{Levin_2012}%
  \BibitemOpen
  \bibfield  {author} {\bibinfo {author} {\bibfnamefont {Y.}~\bibnamefont
  {Levin}},\ }\bibfield  {title} {\bibinfo {title} {Creep events and creep
  noise in gravitational-wave interferometers: Basic formalism and stationary
  limit},\ }\href {https://doi.org/10.1103/PhysRevD.86.122004} {\bibfield
  {journal} {\bibinfo  {journal} {Phys. Rev. D}\ }\textbf {\bibinfo {volume}
  {86}},\ \bibinfo {pages} {122004} (\bibinfo {year} {2012})}\BibitemShut
  {NoStop}%
\bibitem [{\citenamefont {Vajente}(2017)}]{Vajente_2017}%
  \BibitemOpen
  \bibfield  {author} {\bibinfo {author} {\bibfnamefont {G.}~\bibnamefont
  {Vajente}},\ }\bibfield  {title} {\bibinfo {title} {Crackling noise in
  advanced gravitational wave detectors: A model of the steel cantilevers used
  in the test mass suspensions},\ }\href
  {https://doi.org/10.1103/PhysRevD.96.022003} {\bibfield  {journal} {\bibinfo
  {journal} {Phys. Rev. D}\ }\textbf {\bibinfo {volume} {96}},\ \bibinfo
  {pages} {022003} (\bibinfo {year} {2017})}\BibitemShut {NoStop}%
\bibitem [{\citenamefont {Popovi{\'{c} }}\ \emph {et~al.}(2022)\citenamefont
  {Popovi{\'{c} }}, \citenamefont {de~Geus}, \citenamefont {Ji}, \citenamefont
  {Rosso},\ and\ \citenamefont {Wyart}}]{Popovi_2022}%
  \BibitemOpen
  \bibfield  {author} {\bibinfo {author} {\bibfnamefont {M.}~\bibnamefont
  {Popovi{\'{c} }}}, \bibinfo {author} {\bibfnamefont {T.~W.}\ \bibnamefont
  {de~Geus}}, \bibinfo {author} {\bibfnamefont {W.}~\bibnamefont {Ji}},
  \bibinfo {author} {\bibfnamefont {A.}~\bibnamefont {Rosso}},\ and\ \bibinfo
  {author} {\bibfnamefont {M.}~\bibnamefont {Wyart}},\ }\bibfield  {title}
  {\bibinfo {title} {Scaling description of creep flow in amorphous solids},\
  }\bibfield  {journal} {\bibinfo  {journal} {Physical Review Letters}\
  }\textbf {\bibinfo {volume} {129}},\ \href
  {https://doi.org/10.1103/physrevlett.129.208001}
  {10.1103/physrevlett.129.208001} (\bibinfo {year} {2022})\BibitemShut
  {NoStop}%
\bibitem [{\citenamefont {Cooper}\ \emph {et~al.}(2023)\citenamefont {Cooper},
  \citenamefont {Mow-Lowry}, \citenamefont {Hoyland}, \citenamefont {Bryant},
  \citenamefont {Ubhi}, \citenamefont {O’Dell}, \citenamefont {Huddart},
  \citenamefont {Aston},\ and\ \citenamefont {Vecchio}}]{Cooper2022_BOSEM}%
  \BibitemOpen
  \bibfield  {author} {\bibinfo {author} {\bibfnamefont {S.~J.}\ \bibnamefont
  {Cooper}}, \bibinfo {author} {\bibfnamefont {C.~M.}\ \bibnamefont
  {Mow-Lowry}}, \bibinfo {author} {\bibfnamefont {D.}~\bibnamefont {Hoyland}},
  \bibinfo {author} {\bibfnamefont {J.}~\bibnamefont {Bryant}}, \bibinfo
  {author} {\bibfnamefont {A.}~\bibnamefont {Ubhi}}, \bibinfo {author}
  {\bibfnamefont {J.}~\bibnamefont {O’Dell}}, \bibinfo {author}
  {\bibfnamefont {A.}~\bibnamefont {Huddart}}, \bibinfo {author} {\bibfnamefont
  {S.}~\bibnamefont {Aston}},\ and\ \bibinfo {author} {\bibfnamefont
  {A.}~\bibnamefont {Vecchio}},\ }\bibfield  {title} {\bibinfo {title} {Sensors
  and actuators for the advanced ligo a+ upgrade},\ }\href
  {https://doi.org/10.1063/5.0117605} {\bibfield  {journal} {\bibinfo
  {journal} {Review of Scientific Instruments}\ }\textbf {\bibinfo {volume}
  {94}},\ \bibinfo {pages} {014502} (\bibinfo {year} {2023})},\ \Eprint
  {https://arxiv.org/abs/https://doi.org/10.1063/5.0117605}
  {https://doi.org/10.1063/5.0117605} \BibitemShut {NoStop}%
\bibitem [{\citenamefont {Ubhi}\ \emph
  {et~al.}(2022{\natexlab{c}})\citenamefont {Ubhi}, \citenamefont {Bryant},
  \citenamefont {Hoyland},\ and\ \citenamefont {Martynov}}]{Ubhi_2022c}%
  \BibitemOpen
  \bibfield  {author} {\bibinfo {author} {\bibfnamefont {A.~S.}\ \bibnamefont
  {Ubhi}}, \bibinfo {author} {\bibfnamefont {J.}~\bibnamefont {Bryant}},
  \bibinfo {author} {\bibfnamefont {D.}~\bibnamefont {Hoyland}},\ and\ \bibinfo
  {author} {\bibfnamefont {D.}~\bibnamefont {Martynov}},\ }\bibfield  {title}
  {\bibinfo {title} {Cryogenic optical shadow sensors for gravitational wave
  detectors},\ }\href
  {https://doi.org/https://doi.org/10.1016/j.cryogenics.2022.103547} {\bibfield
   {journal} {\bibinfo  {journal} {Cryogenics}\ }\textbf {\bibinfo {volume}
  {126}},\ \bibinfo {pages} {103547} (\bibinfo {year}
  {2022}{\natexlab{c}})}\BibitemShut {NoStop}%
\bibitem [{\citenamefont {Strain}\ and\ \citenamefont
  {Shapiro}(2012)}]{Strain2012}%
  \BibitemOpen
  \bibfield  {author} {\bibinfo {author} {\bibfnamefont {K.~A.}\ \bibnamefont
  {Strain}}\ and\ \bibinfo {author} {\bibfnamefont {B.~N.}\ \bibnamefont
  {Shapiro}},\ }\bibfield  {title} {\bibinfo {title} {Damping and local control
  of mirror suspensions for laser interferometric gravitational wave
  detectors},\ }\href {https://doi.org/10.1063/1.4704459} {\bibfield  {journal}
  {\bibinfo  {journal} {Rev. Sci. Instrum.}\ }\textbf {\bibinfo {volume}
  {83}},\ \bibinfo {pages} {044501} (\bibinfo {year} {2012})}\BibitemShut
  {NoStop}%
\bibitem [{\citenamefont {Cooper}\ \emph {et~al.}(2018)\citenamefont {Cooper},
  \citenamefont {Collins}, \citenamefont {Green} \emph {et~al.}}]{Cooper_2018}%
  \BibitemOpen
  \bibfield  {author} {\bibinfo {author} {\bibfnamefont {S.~J.}\ \bibnamefont
  {Cooper}}, \bibinfo {author} {\bibfnamefont {C.~J.}\ \bibnamefont {Collins}},
  \bibinfo {author} {\bibfnamefont {A.~C.}\ \bibnamefont {Green}}, \emph
  {et~al.},\ }\bibfield  {title} {\bibinfo {title} {A compact, large-range
  interferometer for precision measurement and inertial sensing},\ }\href
  {https://doi.org/10.1088/1361-6382/aab2e9} {\bibfield  {journal} {\bibinfo
  {journal} {Classical and Quantum Gravity}\ }\textbf {\bibinfo {volume}
  {35}},\ \bibinfo {pages} {095007} (\bibinfo {year} {2018})}\BibitemShut
  {NoStop}%
\bibitem [{\citenamefont {Cooper}\ \emph {et~al.}(2022)\citenamefont {Cooper},
  \citenamefont {Collins}, \citenamefont {Prokhorov}, \citenamefont {Warner},
  \citenamefont {Hoyland},\ and\ \citenamefont {Mow-Lowry}}]{Cooper2022_L4C}%
  \BibitemOpen
  \bibfield  {author} {\bibinfo {author} {\bibfnamefont {S.~J.}\ \bibnamefont
  {Cooper}}, \bibinfo {author} {\bibfnamefont {C.~J.}\ \bibnamefont {Collins}},
  \bibinfo {author} {\bibfnamefont {L.}~\bibnamefont {Prokhorov}}, \bibinfo
  {author} {\bibfnamefont {J.}~\bibnamefont {Warner}}, \bibinfo {author}
  {\bibfnamefont {D.}~\bibnamefont {Hoyland}},\ and\ \bibinfo {author}
  {\bibfnamefont {C.~M.}\ \bibnamefont {Mow-Lowry}},\ }\bibfield  {title}
  {\bibinfo {title} {Interferometric sensing of a commercial geophone},\ }\href
  {https://doi.org/10.1088/1361-6382/ac595c} {\bibfield  {journal} {\bibinfo
  {journal} {Classical and Quantum Gravity}\ }\textbf {\bibinfo {volume}
  {39}},\ \bibinfo {pages} {075023} (\bibinfo {year} {2022})}\BibitemShut
  {NoStop}%
\bibitem [{\citenamefont {Smetana}\ \emph {et~al.}(2022)\citenamefont
  {Smetana}, \citenamefont {Walters}, \citenamefont {Bauchinger} \emph
  {et~al.}}]{Smetana2022compact}%
  \BibitemOpen
  \bibfield  {author} {\bibinfo {author} {\bibfnamefont {J.}~\bibnamefont
  {Smetana}}, \bibinfo {author} {\bibfnamefont {R.}~\bibnamefont {Walters}},
  \bibinfo {author} {\bibfnamefont {S.}~\bibnamefont {Bauchinger}}, \emph
  {et~al.},\ }\bibfield  {title} {\bibinfo {title} {Compact michelson
  interferometers with subpicometer sensitivity},\ }\href
  {https://doi.org/10.1103/PhysRevApplied.18.034040} {\bibfield  {journal}
  {\bibinfo  {journal} {Phys. Rev. Applied}\ }\textbf {\bibinfo {volume}
  {18}},\ \bibinfo {pages} {034040} (\bibinfo {year} {2022})}\BibitemShut
  {NoStop}%
\bibitem [{\citenamefont {Isleif}\ \emph {et~al.}(2019)\citenamefont {Isleif},
  \citenamefont {Heinzel}, \citenamefont {Mehmet},\ and\ \citenamefont
  {Gerberding}}]{Isleif_2019}%
  \BibitemOpen
  \bibfield  {author} {\bibinfo {author} {\bibfnamefont {K.-S.}\ \bibnamefont
  {Isleif}}, \bibinfo {author} {\bibfnamefont {G.}~\bibnamefont {Heinzel}},
  \bibinfo {author} {\bibfnamefont {M.}~\bibnamefont {Mehmet}},\ and\ \bibinfo
  {author} {\bibfnamefont {O.}~\bibnamefont {Gerberding}},\ }\bibfield  {title}
  {\bibinfo {title} {Compact multifringe interferometry with subpicometer
  precision},\ }\bibfield  {journal} {\bibinfo  {journal} {Physical Review
  Applied}\ }\textbf {\bibinfo {volume} {12}},\ \href
  {https://doi.org/10.1103/physrevapplied.12.034025}
  {10.1103/physrevapplied.12.034025} (\bibinfo {year} {2019})\BibitemShut
  {NoStop}%
\bibitem [{\citenamefont {Gerberding}\ and\ \citenamefont
  {Isleif}(2021)}]{Gerberding_2021}%
  \BibitemOpen
  \bibfield  {author} {\bibinfo {author} {\bibfnamefont {O.}~\bibnamefont
  {Gerberding}}\ and\ \bibinfo {author} {\bibfnamefont {K.-S.}\ \bibnamefont
  {Isleif}},\ }\bibfield  {title} {\bibinfo {title} {Ghost beam suppression in
  deep frequency modulation interferometry for compact on-axis optical heads},\
  }\bibfield  {journal} {\bibinfo  {journal} {Sensors}\ }\textbf {\bibinfo
  {volume} {21}},\ \href {https://doi.org/10.3390/s21051708}
  {10.3390/s21051708} (\bibinfo {year} {2021})\BibitemShut {NoStop}%
\bibitem [{\citenamefont {Gerberding}\ \emph {et~al.}(2017)\citenamefont
  {Gerberding}, \citenamefont {Isleif}, \citenamefont {Mehmet} \emph
  {et~al.}}]{Gerberding_2017}%
  \BibitemOpen
  \bibfield  {author} {\bibinfo {author} {\bibfnamefont {O.}~\bibnamefont
  {Gerberding}}, \bibinfo {author} {\bibfnamefont {K.-S.}\ \bibnamefont
  {Isleif}}, \bibinfo {author} {\bibfnamefont {M.}~\bibnamefont {Mehmet}},
  \emph {et~al.},\ }\bibfield  {title} {\bibinfo {title} {Laser-frequency
  stabilization via a quasimonolithic mach-zehnder interferometer with arms of
  unequal length and balanced dc readout},\ }\href
  {https://doi.org/10.1103/PhysRevApplied.7.024027} {\bibfield  {journal}
  {\bibinfo  {journal} {Phys. Rev. Applied}\ }\textbf {\bibinfo {volume} {7}},\
  \bibinfo {pages} {024027} (\bibinfo {year} {2017})}\BibitemShut {NoStop}%
\bibitem [{\citenamefont {Armano}\ \emph {et~al.}(2016)\citenamefont {Armano},
  \citenamefont {Audley}, \citenamefont {Auger}, \citenamefont {Baird},
  \citenamefont {Bassan}, \citenamefont {Binetruy}, \citenamefont {Born},
  \citenamefont {Bortoluzzi}, \citenamefont {Brandt}, \citenamefont {Caleno}
  \emph {et~al.}}]{Armano_2016}%
  \BibitemOpen
  \bibfield  {author} {\bibinfo {author} {\bibfnamefont {M.}~\bibnamefont
  {Armano}}, \bibinfo {author} {\bibfnamefont {H.}~\bibnamefont {Audley}},
  \bibinfo {author} {\bibfnamefont {G.}~\bibnamefont {Auger}}, \bibinfo
  {author} {\bibfnamefont {J.~T.}\ \bibnamefont {Baird}}, \bibinfo {author}
  {\bibfnamefont {M.}~\bibnamefont {Bassan}}, \bibinfo {author} {\bibfnamefont
  {P.}~\bibnamefont {Binetruy}}, \bibinfo {author} {\bibfnamefont
  {M.}~\bibnamefont {Born}}, \bibinfo {author} {\bibfnamefont {D.}~\bibnamefont
  {Bortoluzzi}}, \bibinfo {author} {\bibfnamefont {N.}~\bibnamefont {Brandt}},
  \bibinfo {author} {\bibfnamefont {M.}~\bibnamefont {Caleno}}, \emph
  {et~al.},\ }\bibfield  {title} {\bibinfo {title} {Sub-femto-$g$ free fall for
  space-based gravitational wave observatories: Lisa pathfinder results},\
  }\href {https://doi.org/10.1103/PhysRevLett.116.231101} {\bibfield  {journal}
  {\bibinfo  {journal} {Phys. Rev. Lett.}\ }\textbf {\bibinfo {volume} {116}},\
  \bibinfo {pages} {231101} (\bibinfo {year} {2016})}\BibitemShut {NoStop}%
\bibitem [{\citenamefont {Weichert}\ \emph {et~al.}(2012)\citenamefont
  {Weichert}, \citenamefont {Köchert}, \citenamefont {Köning}, \citenamefont
  {Flügge}, \citenamefont {Andreas}, \citenamefont {Kuetgens},\ and\
  \citenamefont {Yacoot}}]{Weichert_2012}%
  \BibitemOpen
  \bibfield  {author} {\bibinfo {author} {\bibfnamefont {C.}~\bibnamefont
  {Weichert}}, \bibinfo {author} {\bibfnamefont {P.}~\bibnamefont {Köchert}},
  \bibinfo {author} {\bibfnamefont {R.}~\bibnamefont {Köning}}, \bibinfo
  {author} {\bibfnamefont {J.}~\bibnamefont {Flügge}}, \bibinfo {author}
  {\bibfnamefont {B.}~\bibnamefont {Andreas}}, \bibinfo {author} {\bibfnamefont
  {U.}~\bibnamefont {Kuetgens}},\ and\ \bibinfo {author} {\bibfnamefont
  {A.}~\bibnamefont {Yacoot}},\ }\bibfield  {title} {\bibinfo {title} {A
  heterodyne interferometer with periodic nonlinearities smaller than
  $\pm$10{\hspace{0.167em}}pm},\ }\href
  {https://doi.org/10.1088/0957-0233/23/9/094005} {\bibfield  {journal}
  {\bibinfo  {journal} {Measurement Science and Technology}\ }\textbf {\bibinfo
  {volume} {23}},\ \bibinfo {pages} {094005} (\bibinfo {year}
  {2012})}\BibitemShut {NoStop}%
\bibitem [{\citenamefont {de~la Rue}\ \emph {et~al.}(1972)\citenamefont {de~la
  Rue}, \citenamefont {Humphryes}, \citenamefont {Mason},\ and\ \citenamefont
  {Ash}}]{Rue_1972}%
  \BibitemOpen
  \bibfield  {author} {\bibinfo {author} {\bibfnamefont {R.}~\bibnamefont
  {de~la Rue}}, \bibinfo {author} {\bibfnamefont {R.}~\bibnamefont
  {Humphryes}}, \bibinfo {author} {\bibfnamefont {I.}~\bibnamefont {Mason}},\
  and\ \bibinfo {author} {\bibfnamefont {E.}~\bibnamefont {Ash}},\ }\bibfield
  {title} {{\selectlanguage {English}\bibinfo {title} {Acoustic-surface-wave
  amplitude and phase measurements using laser probes}},\ }\href
  {https://digital-library.theiet.org/content/journals/10.1049/piee.1972.0021}
  {\bibfield  {journal} {\bibinfo  {journal} {Proceedings of the Institution of
  Electrical Engineers}\ }\textbf {\bibinfo {volume} {119}},\ \bibinfo {pages}
  {117} (\bibinfo {year} {1972})}\BibitemShut {NoStop}%
\bibitem [{\citenamefont {Heinzel}\ \emph {et~al.}(2010)\citenamefont
  {Heinzel}, \citenamefont {Cervantes}, \citenamefont {Mar\'{i}n},
  \citenamefont {Kullmann}, \citenamefont {Feng},\ and\ \citenamefont
  {Danzmann}}]{Heinzel_2010}%
  \BibitemOpen
  \bibfield  {author} {\bibinfo {author} {\bibfnamefont {G.}~\bibnamefont
  {Heinzel}}, \bibinfo {author} {\bibfnamefont {F.~G.}\ \bibnamefont
  {Cervantes}}, \bibinfo {author} {\bibfnamefont {A.~F.~G.}\ \bibnamefont
  {Mar\'{i}n}}, \bibinfo {author} {\bibfnamefont {J.}~\bibnamefont {Kullmann}},
  \bibinfo {author} {\bibfnamefont {W.}~\bibnamefont {Feng}},\ and\ \bibinfo
  {author} {\bibfnamefont {K.}~\bibnamefont {Danzmann}},\ }\bibfield  {title}
  {\bibinfo {title} {Deep phase modulation interferometry},\ }\href
  {https://doi.org/10.1364/OE.18.019076} {\bibfield  {journal} {\bibinfo
  {journal} {Opt. Express}\ }\textbf {\bibinfo {volume} {18}},\ \bibinfo
  {pages} {19076} (\bibinfo {year} {2010})}\BibitemShut {NoStop}%
\end{thebibliography}%

\end{document}